\documentclass{article}
\usepackage{graphicx}
\usepackage{amsmath}

\input{tcilatex}

\begin{document}

\title{The ideal trefoil knot}
\author{P. Pieranski and S. Przybyl \\
Poznan University of Technology\\
Nieszawska 13A, 60 965 Poznan\\
Poland\\
e-mail: Piotr.Pieranski@put.poznan.pl}
\maketitle

\begin{abstract}
The most tight conformation of the trefoil knot found by the SONO algorithm
is presented. Structure of the set of its self-contact points is analyzed.
\end{abstract}

\section{Introduction}

Finding the best way of packing a tube within a box seems to be rather\ a
gardening than a scientific problem. However, the optimal single helix,
discovered in a computer simulation study of this problem, \cite{Maritan}
and \cite{SPPPhelices}, proves to be ubiquitous in many proteins as their $%
\alpha $-helical parts. It seems, as suggested in \cite{Stasiak & Maddocks},
that also the closely packed double helix appearing in the process of
twisting two ropes together \cite{Skretka} have been already discovered and
applied by nature. Laboratory experiments allow one to observe in the real
time how \ the optimal helices are formed in various systems e.g. the
bacterial flagellas \cite{Flagella} or phospholipid membranes \cite
{membranes}.

Both processes, of packing the ropes and twisting them together, occur
simultaneously when a knot tied on a rope becomes tightened. The problem of
finding the most tight, least rope consuming conformations of knots was
independently posed and indicated as essential by different authors; for
references see \cite{Ideal knots}. Knots in such optimal, most tight
conformations are often called \textit{ideal}, a term proposed by Simon \cite
{Simon96}, and introduced into the literature by Stasiak \cite{Katritch_1}.
Ideal conformations minimize the value of the size-invariant variable $%
\Lambda =L/D$, where $L$ and $D$ are, respectively, the length and the
diameter of the \textit{perfect rope }(defined below) on which the knot is
tied. The only knot whose ideal conformation is known at present is the
trivial knot (unknot). See Fig.\ref{Fig1}. \FRAME{ftbpFU}{1.9579in}{1.5956in%
}{0pt}{\Qcb{Ideal unknot.}}{\Qlb{Fig1}}{idealunknot.jpg}{\special{language
"Scientific Word";type "GRAPHIC";maintain-aspect-ratio TRUE;display
"USEDEF";valid_file "F";width 1.9579in;height 1.5956in;depth
0pt;original-width 7.4478in;original-height 6.0623in;cropleft "0";croptop
"1";cropright "1";cropbottom "0";filename '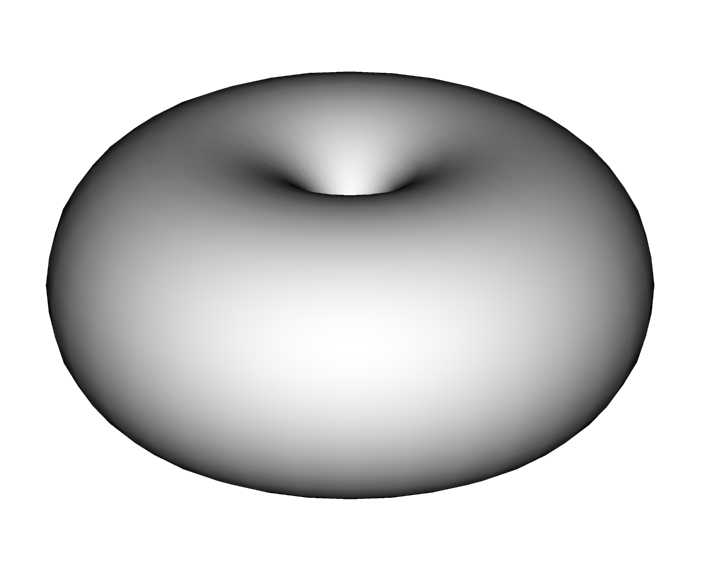';file-properties
"XNPEU";}}Its length in the ideal, circular conformation equals $\pi D$,
thus $\Lambda =\pi $. Finding the ideal conformation of a nontrivial knot is
a nontrivial task. Initiated a few years ago search for the ideal
conformations of nontrivial knots continues.

One of the algorithms used in the search is SONO (Shrink-On-No-Overlaps)\cite
{ProDialog1}. SONO simulates a process in which the rope, on which a knot is
tied, slowly shrinks. The rope is allowed to shrink only when no overlaps of
the rope with itself are detected within the knot. When such overlaps occur,
SONO modifies the knot conformation to remove them. If this is no more
possible, the process ends. Unfortunately, ending of the tightening process
does not mean that the ideal conformation of a given knot was found. The
tightening process could have stopped also because a local minimum of the
thickness energy was entered. The possibility that there exists a different,
less rope consuming conformation, cannot be excluded.

SONO has been used in the search of ideal conformations of both prime and
composite knots. Parameters of the least rope consuming conformations found
by the algorithm were listed in \cite{Katritch_2} and \cite{My chapter}. In
a few cases, SONO managed to find better conformations than the simulated
annealing procedure \cite{Katritch_1}. However, for the most simple knots,
in particular, the trefoil knot, the simulated annealing and SONO provided
identical results; the $\Lambda $ values are identical within experimental
errors. It seems obvious, that no better conformations of the knot exist. We
feel obliged to emphasize, however, that it is only an intuitively obvious
conclusion - no formal proofs have been provided so far. As indicated in 
\cite{Stasiak & Maddocks}, we are in a situation similar to that, which
lasted in the problem of the best packing of spheres for 400 years. That the
face centered cubic and hexagonal close packed lattices were among the
structures which minimize the volume occupied by closely packed hard spheres
seemed to be obvious since the times of Kepler, however the formal proof of
the conjecture was provided but a few years ago \cite{Sloane}. Waiting for
the formal proofs that what we have observed in the knot tightening
numerical experiments is the ideal conformation of the trefoil, seems to be
a too cautious attitude. Thus, after a few years of experimenting, we
decided to present the best, least rope consuming conformation of the
trefoil knot we managed to find. We compare it with the most tight
conformation of the knot which can be found within the analytically defined
family of torus knots. In particular, we describe the qualitative change in
the set of self-contacts which takes place within the trefoil knot during
the tightening process. We believe that some of the features of the
self-contact set we have found may be present also in ideal conformations of
other knot types.

An alternative method of searching for the most tight conformations of knots
consists in inflating the rope on which the knot has been tied In such a
process the length of the rope is kept fixed. The maximum radius to which
the rope in a given conformation of a knot can be inflated is closely
related with the injectivity radius considered in detail by Rawdon \cite
{Rawdon}.

\section{The perfect rope}

It is the aim of the computer simulations we perform to simulate the
tightening process of knots tied on the \textit{perfect rope}: perfectly
flexible, but at the same time perfectly hard in its circular cross-section.
The surface of the perfect rope can be seen as the union of all circles
centered on and perpendicular to the knot axis $C$. See Fig.\ref
{perfect_rope}.\FRAME{ftbFU}{2.904in}{1.7936in}{0pt}{\Qcb{The perfect rope.
Perpendicular sections of the rope are of the disk shape. None of the disks
are allowed to overlap. This puts a limit not only on the spacial distance
of different fragments of the curve into which the rope is shaped, but also
on its local curvature.}}{\Qlb{perfect_rope}}{helice2.jpg}{\special{language
"Scientific Word";type "GRAPHIC";maintain-aspect-ratio TRUE;display
"USEDEF";valid_file "F";width 2.904in;height 1.7936in;depth
0pt;original-width 9.052in;original-height 5.5728in;cropleft "0";croptop
"1";cropright "1";cropbottom "0";filename 'Helice2.jpg';file-properties
"XNPEU";}}

We assume that $C$ is smooth and simple, i.e. self-avoiding, what guaranties
that at each of its points $\mathbf{r}$ the tangent vectors $\mathbf{\tau }(%
\mathbf{r}),$ and thus the circular cross-section, are well defined. The
surface remains smooth as long as:

A. the \textit{local curvature radius} $r_{\kappa }$ of the knot axis is
nowhere smaller than $D/2$,

B. the \textit{minimum distance of closest approach }$d_{\ast }$ is nowhere
smaller then $D/2$ .

The \textit{minimum distance of closest approach }$d_{\ast }$, known also as
the \textit{doubly critical self-distance}, see \cite{Simon96}, is defined
in \cite{Gonzalez}, as the smallest distance between all pairs of points $(%
\mathbf{r}_{1},\mathbf{r}_{2})$ on the knot axis, having the property, that
the vector $(\mathbf{r}_{2}-\mathbf{r}_{1})$ joining them is orthogonal to
the tangent vectors $\mathbf{\tau }(\mathbf{r}_{1}),\mathbf{\tau }(\mathbf{r}%
_{2})$ located at the points: 
\begin{equation}
d_{\ast }(C)=\underset{\mathbf{r}_{1},\mathbf{r}_{2}\in C}{\min }\left\{ |%
\mathbf{r}_{2}-\mathbf{r}_{1}|:\mathbf{\tau }(\mathbf{r}_{1})\perp (\mathbf{r%
}_{2}-\mathbf{r}_{1}),\mathbf{\tau }(\mathbf{r}_{2})\perp (\mathbf{r}_{2}-%
\mathbf{r}_{1})\right\}  \label{eq4}
\end{equation}

As shown by Gonzalez and Maddocks \cite{Gonzalez}, the two conditions can be
gathered into a single one providing that the notion of the \textit{global
curvature radius} $\rho _{G}$ is introduced: 
\begin{equation}
\rho _{G}(\mathbf{r}_{1})=\underset{\mathbf{r}_{1}\neq \mathbf{r}_{2}\neq 
\mathbf{r}_{3}\neq \mathbf{r}_{1}}{\underset{\mathbf{r}_{2},\mathbf{r}%
_{3}\in C}{\min }}\rho (\mathbf{r}_{1},\mathbf{r}_{2},\mathbf{r}_{3})
\label{eq5}
\end{equation}
where, $\rho (\mathbf{r}_{1},\mathbf{r}_{2},\mathbf{r}_{3})$ is the radius
of the unique circle (the circumcircle) which passes through all of the
three points: $\mathbf{r}_{1},\mathbf{r}_{2}$ and $\mathbf{r}_{3}$. Using
the notion of the global curvature, the condition which guaranties
smoothness of the knot surface can be reformulated as follows:

C. the global curvature radius $\rho _{G}$ of the knot axis is nowhere
smaller than $D/2$.

Analysis of the conformations produced by the SONO\ algorithm proves that
conditions A and B, (and C) are fulfilled.

\section{Parametrically tied trefoil knot}

The trefoil knot can be tied on the surface of a torus. See Fig.\ref{31torus}%
\FRAME{ftbFU}{4.0335in}{2.4803in}{0pt}{\Qcb{The trefoil knot is a torus knot
- it can be tied on the surface of a torus.}}{\Qlb{31torus}}{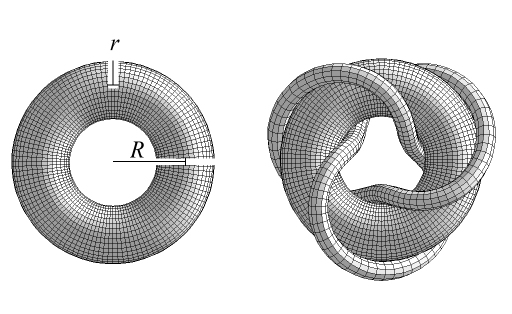}{%
\special{language "Scientific Word";type "GRAPHIC";maintain-aspect-ratio
TRUE;display "USEDEF";valid_file "F";width 4.0335in;height 2.4803in;depth
0pt;original-width 10.6666in;original-height 6.5414in;cropleft "0";croptop
"1";cropright "1";cropbottom "0";filename '31torus.jpg';file-properties
"XNPEU";}}Consider the set of 3 periodic functions: 
\begin{eqnarray}
x &=&[R+r\cos (2\,\nu _{1}\,\pi \,t)]\,sin(2\,\nu _{2}\,\pi \,t)
\label{eq1a} \\
y &=&[R+r\cos (2\,\nu _{1}\,\pi \,t)]\,cos(2\,\nu _{2}\,\pi \,t)\qquad 
\label{eq1b} \\
z &=&r\sin (2\,\nu _{1}\,\pi \,t)  \label{eq1c}
\end{eqnarray}

The trajectory determined by equations \ref{eq1a}, \ref{eq1b} and \ref{eq1c}
becomes closed as $t$ spans a unit interval. For the sake of simplicity we
shall consider the $[0,1)$ interval. For all relatively prime integer values
of $\nu _{1}$, $\nu _{2}$ equations \ref{eq1a}, \ref{eq1b} and \ref{eq1c}
define self-avoiding closed curves located on the surface of a torus. $R$
denotes here the radius of the circle determining the central axis of the
torus while $r$ denotes the radius of its circular cross-sections. For the
trefoil knot, frequencies $\nu _{1}$, $\nu _{2}$ equal $2$ and $3$,
respectively. In what follows we consider knots tied on a rope; trajectories
defined by equations \ref{eq1a}, \ref{eq1b} and \ref{eq1c} determine
position of its axis.

The $(\nu _{1},\nu _{2})$ and the $(\nu _{2},\nu _{1})$ torus knots are
ambient isotopic, i.e. they can be transformed one into another without
cutting the rope on which they are tied \cite{Knot Book}. As shown
previously, the $(2,3)$ version of the trefoil is less rope consuming \cite
{My chapter}. Thus, the $(3,2)$ version will not be discussed below.

Assume that the trefoil knot whose axis is defined by equations \ref{eq1a}, 
\ref{eq1b} and \ref{eq1c} is tied on a rope of diameter $D=1$. In what
follows we shall refer to it as the \textit{parametrically tied trefoil }%
(PTT)\textit{\ knot}. In such a case, radius $r$ of the torus on which the
axis of knot is located, cannot be smaller than $1/2$ ; below this value
overlaps of the rope with itself will certainly appear; at $r=1/2$ the rope
remains in a continuous self-contact along the torus axis. To keep the
self-contacts we assume in what follows that $r=1/2$. To check, if the knot
is free of overlaps in other regions, one can analyze the map of its
internal distances. Let $t_{1}$ and $t_{2}$ be two values of the parameter $%
t $, both located in the $[0,1)$ interval. Let $(x_{1},y_{1},z_{1})$ and $%
(x_{2},y_{2},z_{2})$ be the coordinates of two points indicated within the
knot axis by $t_{1}$ and $t_{2}$, respectively. Let $d(t_{1},t_{2})$ be the
Euclidean distance between the points: 
\begin{equation}
d(t_{1},t_{2})=\sqrt{(x_{2}-x_{1})^{2}+(y_{2}-y_{1})^{2}+(z_{2}-z_{1})^{2}}
\label{eq6}
\end{equation}

\ The map of the function, see Fig.\ref{31interdistance_panoramic} displays
a mirror symmetry induced by the equality $d(t_{1},t_{2})=d(t_{2},t_{1})$.

\FRAME{ftbFU}{4.3984in}{2.7043in}{0pt}{\Qcb{The map of the intraknot
distances of the most tight PPT knot.}}{\Qlb{31interdistance_panoramic}}{%
fig3_11.jpg}{\special{language "Scientific Word";type
"GRAPHIC";maintain-aspect-ratio TRUE;display "USEDEF";valid_file "F";width
4.3984in;height 2.7043in;depth 0pt;original-width 10.6666in;original-height
6.5414in;cropleft "0";croptop "1";cropright "1";cropbottom "0";filename
'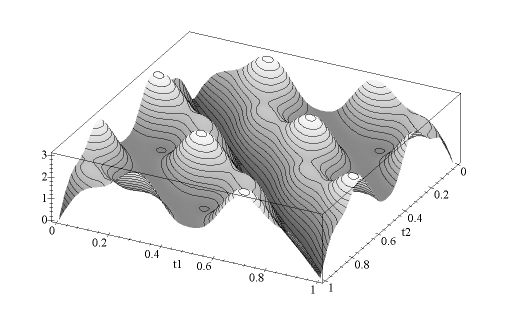';file-properties "XNPEU";}}

Looking for possible overlaps within the knot one looks for regions within
the internal distances landscape, where $d(t_{1},t_{2})<1$. The most visible
depression within the landscape of the interknot distances is located around
the diagonal where $t_{1}=t_{2}$. As easy to see, $d(t_{1},t_{2})=0$ \ along
the line, but for obvious reasons this does not implies any overlaps within
the knot.

Another valley within which $d(t_{1},t_{2})$ may go down to the critical $1$
value is localized in the vicinity of lines defined by equality $%
|t_{2}-t_{1}|=1/2$. To see, if in the vicinity of the lines the height
really drops to or even below 1, we plotted the map of the $d(t_{1},t_{2})$
function in such a manner, that regions lying below the arbitrarily chosen $%
1.005$ level were cut off.

\bigskip \FRAME{ftbFU}{4.1866in}{3.9479in}{0pt}{\Qcb{The map of the
intraknot distances. Left - the most tight PTT knot. Right - the most tight
STT knot. The map was cut from below at the height 10.005. }}{\Qlb{Fig4}}{%
fig4_1005.jpg}{\special{language "Scientific Word";type
"GRAPHIC";maintain-aspect-ratio TRUE;display "USEDEF";valid_file "F";width
4.1866in;height 3.9479in;depth 0pt;original-width 6.5311in;original-height
6.1566in;cropleft "0";croptop "1";cropright "1";cropbottom "0";filename
'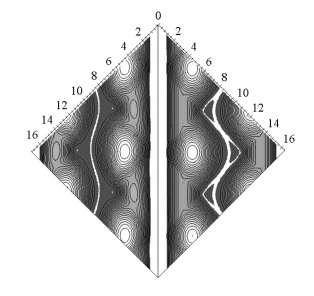';file-properties "XNPEU";}}

As seen in Fig.\ref{Fig4} there are four such regions within the PTT knot:
one in the shape of a sinusoidal band and three in shapes of almost circular
patches. The band contains in its middle the mentioned above continuous line
of self-contacts points; it is the axis of the torus on which the knot is
tied. The circular patches contain 3 additional contact points; when $R$
becomes too small, overlaps appear around the points. Numerical analysis we
performed reveals that (with the 5 decimal digits accuracy we applied) the
overlaps occurring within these regions vanish above $R=1.1158$. For $%
R=1.1159$ the distance between the closest points located within these
regions of the knot equals $0.9999$. For $R=1.1158$ the distance is equal $%
1.0000$. Where, within the PTT knot the self-contact points are located is
shown in Fig.\ref{Fig5}

\FRAME{ftbFU}{3.1107in}{2.9257in}{0pt}{\Qcb{Localization of the set of
self-contact points within the most tight PPT knot.}}{\Qlb{Fig5}}{fig5.jpg}{%
\special{language "Scientific Word";type "GRAPHIC";maintain-aspect-ratio
TRUE;display "USEDEF";valid_file "F";width 3.1107in;height 2.9257in;depth
0pt;original-width 5.6982in;original-height 5.354in;cropleft "0";croptop
"1";cropright "1";cropbottom "0";filename '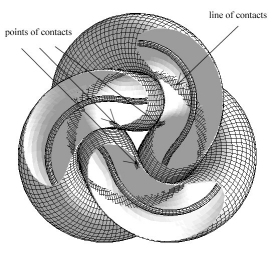';file-properties
"XNPEU";}}

\section{SONO tied trefoil knot}

Considerations presented above indicated the value of $R$, at which the PTT
knot reaches its most tight conformation. The length $L_{t}$ of the rope
engaged in this conformation of the trefoil knot equals $17.0883$. Can one
tie the trefoil knot using a shorter piece of the rope? Theoretical
considerations indicate that this possibility cannot be excluded. As proven
in \cite{Cantarella} the piece of rope used to tie the trefoil knot cannot
be shorter than $Lm=(2+\sqrt{2})\pi \approx 10.72$. Such a location of this
lower limit leaves a lot of place for a possible further tightening of the
knot. Application of SONO reveals that the tightening is possible providing
the conformation of the knot is allowed to leave the subspace of the
parametrically tied torus conformations. This happens spontaneously in
numerical simulations in which the most tight PTT knot is supplied to SONO
as the initial conformation. SONO algorithm manages to make it shorter. In
the simulations we performed, SONO reduced the length of the knot by about
4\% to $L_{exp}=16.38$. The discrete representation of the knot used in the
simulations contained $N=327$ nodes. Below we describe the final
conformation. For the sake of simplicity we shall refer to trefoil knots
processed by the SONO algorithm as the SONO\ tied trefoil (STT) knots.

The differences in the conformation of the most tight conformations of the
PTT and STT knots is a subtle one. The essential difference lies in the
structure of the sets of their self-contact points. As mentioned above, the
circular line of self-contact points present in the family of the PTT knots
stays intact as $R$ is changed within the family. Tightening of a PTT knot
achieved by decreasing the radius $R$ of the torus stops when additional
discrete points of contacts appear at three locations within the knot. This
happens as $R$ becomes equal $1.1158$. Further tightening of the knot within
the family of PTT knots is not possible, it becomes possible within the
family of the STT knots.

\FRAME{ftbFU}{3.3321in}{3.2206in}{0pt}{\Qcb{The set of the self-contact
points in the most tight STT knot as seen within the map of the intraknot
distances.}}{\Qlb{Fig6}}{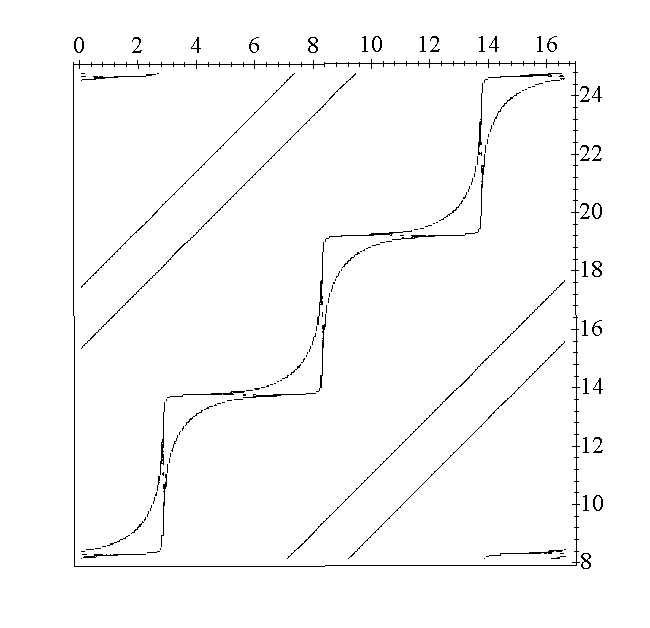}{\special{language "Scientific
Word";type "GRAPHIC";maintain-aspect-ratio TRUE;display "USEDEF";valid_file
"F";width 3.3321in;height 3.2206in;depth 0pt;original-width
6.7706in;original-height 6.5414in;cropleft "0";croptop "1";cropright
"1";cropbottom "0";filename 'map100002.jpg';file-properties "XNPEU";}}

During the tightening process carried out by SONO, the set of the
self-contact points undergoes both qualitative and quantitative changes.
First of all, the line of contacts present in the PTT knot changes its shape
becoming distinctly non-circular. Secondly, the three contact points give
birth to pieces of new line of self-contacts. Unexpectedly, the new pieces
do not connect into a new line, wiggling around and crossing the old line,
but they are mounted into the old line in such a manner, that a single,
self-avoiding and knotted line of self-contacts is created. That this is the
case was revealed by a precise analysis of the interknot distances function.
A map covering the interknot distances only within the very thin $%
[1.00000,1.00002]$ interval shows two separated lines, see Fig.\ref{Fig6},
corresponding to a single, self-avoiding and knotted line of contact.

\FRAME{ftbFU}{2.8764in}{2.6247in}{0pt}{\Qcb{Position of the line of the
self-contact points within the ideal trefoil knot. To make the line more
visible, a part of the knot was cut out.}}{\Qlb{DoubleLineInTheKnot}}{%
doublelineinknotgrey.jpg}{\special{language "Scientific Word";type
"GRAPHIC";maintain-aspect-ratio TRUE;display "USEDEF";valid_file "F";width
2.8764in;height 2.6247in;depth 0pt;original-width 6.6461in;original-height
6.0623in;cropleft "0";croptop "1";cropright "1";cropbottom "0";filename
'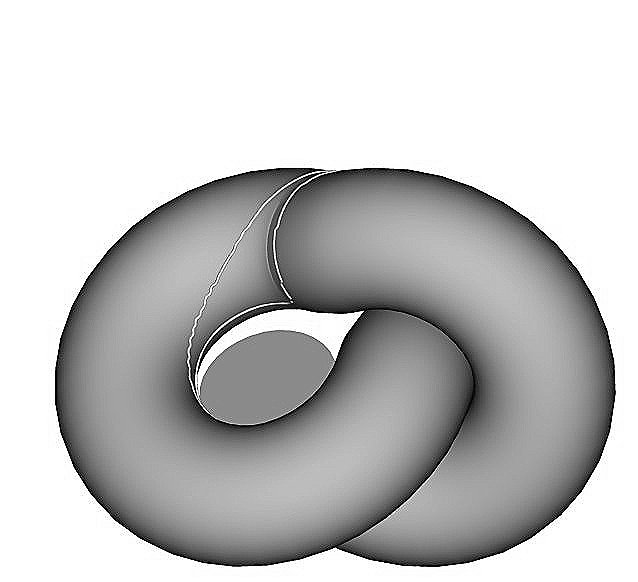';file-properties "XNPEU";}}

In addition to the line, a set of three points of self-contacts is formed.
The points are located at places where the line of self-contacts becomes
almost tangent to itself. The self-contact line runs twice around the knot.
As a result, each of the circular cross-sections of the rope stays here in
touch with another two such sections. The close packed structure formed in
such a manner is much more stable than the structure of the most tight PTT
knot, where single contacts were predominant. Let us note, that figure 1e
presented in ref. \cite{Gonzalez} a similar self-contact line structure can
be seen. Unfortunately, inspecting the figure one cannot see, if the
''self-contact spikes'' shown there form a single, self-avoiding, knotted or
a double, crossing itself line. The problem was not discussed in the text.
Let us emphasize, however, that the difference between the two possibilities
is confined to a zero-measure set.

\section{Discussion}

Ideal knots are objects of which very little is known still. The only knot
whose ideal conformation is known rigorously is the unknot. Its ideal
conformation, a circle of a radius identical with the radius of the rope on
which it is tied, can be conveniently described parametrically. The set of
the self-contact points is here limited to a single point: the center of the
circle. All circular sections of the rope meet at this point. The maximum
local curvature and the minimum double critical self-distance limiting
conditions are simultaneously met.

The situation in the case of the trefoil knot, the simplest non-trivial
prime knot, is radically different. Here the most tight parametrically
defined conformation proves to be not ideal. As demonstrated by the present
authors, it can be tightened more with the use of the SONO\ algorithm. The
set of the self-contact points becomes rebuilt during the tightening
process. Its topology becomes different. In the case the PPT knot the set of
the self-contact points consists of acircle and 3 separated points. As the
numerical experiments we performed suggest, in the case of the STT knot, the
set of the self-contact points turns unexpectedly into a single line. Which
the structure of the set of self contact points in other prime knots is,
remains an open question.

\bigskip

\textbf{Acknowledgment }PP thanks Andrzej Stasiak, John Maddocs, Robert
Kusner, Kenneth Millet, Jason Cantarella and Eric Rawdon for helpful
discussions. This work was carried out under Project KBN 5 PO3B 01220.

\end{document}